# A search on the Klein-Gordon equation

**B. Gönül**

Department of Engineering Physics, University of Gaziantep, 27310, Gaziantep-Türkiye

**Abstract**

The $s$–wave Klein-Gordon equation for the bound states is separated in two parts to see clearly the relativistic contributions to the solution in the non-relativistic limit. The reliability of the model is discussed with the specifically chosen two examples.

With the confidence gained by the successful applications of a novel approach developed recently, see [1] and the related references therein, in solving distinct problems within the frame of non-relativistic physics through the Schrödinger equation, we turn our attention via this Letter to a further application of the model in the relativistic physics hoping to see the power of the formalism leading to the completeness.

Due to the significance of exactly solvable relativistic equations for the systems under the inflence of strong potentials in the vast area of physics, a considerable incresing interest in the study of the Klein-Gordon (K-G) and Dirac equations has appeared in the literature, for a recent review see [2-5]. However, to our knowledge the relation between the strengths of the vector and scaler potentials and the relativistic corrections coming to the non-relativistic solutions has not been fully explored, although the literature involves many valuable applications on this matter.

Within this context, and using the spirit of the work in [1], the work presented in this Letter deals with the K-G equation which is carefully decomposed in two pieces to see unambiguously the behaviour of the spinless particle in the non-relativistic domain and the modifications brought by the relativistic effects. Before proceeding, one needs to visualize the possible non-relativistic limit of this equation through a crude approximation, which will provide a better undertanding in building the formalism presented here.

In the presence of vector and scalar potentials the (1+1)-dimensional time-independent K-G equation for a spinless particle of rest mass $m$ reads

$$-(\hbar c)^2 \psi_n'' + (mc^2 + S)^2 \psi_n = (E_n - V)^2 \psi_n \quad , \qquad n = 0,1,2,... \tag{1}$$

where $E$ is the relativistic energy of the particle, $c$ is the velocity of the light and $\hbar$ is the Planck constant. The vector and scalar potentials are given by $V(r)$ and $S(r)$, respectively. In the non-relativistic approximation (potential energies small compared to $mc^2$ and $E \cong mc^2$), Eq. (1) becomes

$$-\frac{\hbar^2}{2m}\psi_n'' + (S+V)\psi_n \approx (E_n - mc^2)\psi_n \quad . \tag{2}$$

Eq. (2) shows that $\psi$ obeys the Schrödinger equation with binding energy equal to $E - mc^2$, and without distinquishing the contributions of vector and scalar potentials.

Now, bearing in mind this form of the K-G equation in the non-relativistic domain, we suggest that the full relativistic wave function in (1) may be expressed as $\psi = \chi\phi$ where $\chi$ denotes the behaviour of the wave function in the non-relativistic region and $\phi$ is the modification function due to the relativistic effects. This consideration transforms Eq. (1) into a couple of equation

$$\frac{\chi_n''}{\chi_n} = 2(mS + E_n V) - \varepsilon_n \quad , \tag{3}$$

$$\frac{\phi_n''}{\phi_n} + 2\frac{\chi_n'\phi_n'}{\chi_n\phi_n} = (S^2 - V^2) - \Delta\varepsilon_n \quad , \tag{4}$$

where the natural units $\hbar = c = 1$ is employed for a clear comparison with the other works in the literature. In the above equations, $\varepsilon$ and $\Delta\varepsilon$ represent the binding energy within the non-relativistic limit and the modification term because of the relativistic consideration (if any), respectively. Note that $E^2 - m^2 = \varepsilon + \Delta\varepsilon$ and the relativistic corrections are involved within the frame of Eq. (4) in a non-perturbative way. This simple but more flexible presentation of the K-G formalism is compatible with the crude approximation, Eq. (2) used for revealing the appearance of K-G equation in the non-relativistic limit, and also confirms the nice discussion in Ref. [2] about the possible misinterpretation in the related literature regarding the relativistic extensions of the given potentials which behave in a similar manner at the non-relativistic domain. Additionally, it is noticeable that relativistic contributions in case $S = \pm V$ disappear whereas they can be calculated explicitly through (4), which will be discussed below by the examples. Eq (3), yields the free particle solution if $S = -V$ because $E \approx m$ in the limit where this equation is valid while (3) reproduces Shrödinger like non-relativistic solutions for the case $S = V$, which overall justify the reliability of the formalism when the ongoing discussions considered in the literature, e.g. [2, 4].

For practical calculations, Eqs. (3) and (4) are expressed by the Riccati equation,

$$W_n^2 - \frac{W_n'}{\sqrt{2m}} = 2(mS + E_n V) - \varepsilon_n \quad , \quad W_n = -\frac{1}{\sqrt{2m}} \frac{\chi_n'}{\chi_n} \quad , \tag{5}$$

$$\Delta W_n^2 - \frac{\Delta W_n'}{\sqrt{2m}} + 2 W_n \Delta W_n = (S^2 - V^2) - \Delta \varepsilon_n \quad , \quad \Delta W_n = -\frac{1}{\sqrt{2m}} \frac{\phi_n'}{\phi_n} \quad . \tag{6}$$

It is worth to note that if the whole potential $(2mS + 2mV + S^2 - V^2)$ is an exactly solvable then the above equations reduce to a simple form within the framework of the usual supersymmetric quantum theory [6] where a unified treatment like Eq. (5) is employed with $W_n^{SUSY} = W_n + \Delta W_n$. However, if Eq. (6) has no analytical solution one cannot use $W_n^{SUSY}$ concept in dealing with such problems. To overcome this drawback of the formalism, an elegant reliable technique leading to approximate solutions of (6) has been recently introduced in Ref. [1] for any state of interest. Therefore, the standard treatment of the supersymmetric quantum mechanics may be seen as a particular case of the present scheme.

As an illustrative example, we start with the well known Hulthen potential which is frequently used in the literature to justify theoretical models introduced. Considering the related works [5,7], the scaler and vector potentials are chosen as

$$S(r) = -\frac{S_0}{e^{\alpha r} - 1} \quad , \quad V(r) = -\frac{V_0}{e^{\alpha r} - 1} \quad , \tag{7}$$

which, in the light of Eq. (5), restricts us to define

$$W_{n=0} = -\frac{\alpha/\sqrt{2m}}{e^{\alpha r} - 1} + A \quad , \tag{8}$$

leading to $A = \frac{\sqrt{m/2}}{\alpha}\left(U_0 - \frac{\alpha^2}{2m}\right)$ where $U_0 = 2(mS_0 + E_{n=0} V_0)$. The corresponding non-relativistic energy and unnormalized wave function in the ground state

$$\varepsilon_{n=0} = -A^2 = -\frac{(2mU_0 - \alpha^2)^2}{8m\alpha^2} \quad , \quad \chi_{n=0} = e^{-\sqrt{2m}\int W_{n=0} dz} = (1 - e^{-\alpha r}) e^{-\sqrt{2m} A r} \quad , \tag{9}$$

which are in agreement with the work [8] performed in the non-relativistic frame. From Eq. (4), bound states requirements such that $S \succ V$ and $E^2 - m^2 = \varepsilon + \Delta\varepsilon \prec 0$ subsequently $m \succ E$ are satisfied. In this case, with the consideration of (6), we set $\Delta W$

$$\Delta W = -\frac{\delta(\alpha/\sqrt{2m})}{e^{\alpha r} - 1} + B \quad , \tag{10}$$

from where $B = -\sqrt{m/2}\delta U_0/\alpha(\delta+1)$. It is stressed that for $\delta \to 0$, the relativistic effects due to strong interactions die away because $\Delta W \to 0$, together with $\Delta V \to 0$ and $\Delta\varepsilon \to 0$. From equations (6) and (10), in case $\delta \succ 0$, the relativistic contributions to the non-relativistic solutions are

$$\Delta\varepsilon_{n=0} = -B(B + 2A) = \frac{\delta U_0}{2\alpha^2(\delta+1)^2}\left[mU_0(\delta+2) - \alpha^2(\delta+1)\right] \quad ,$$

$$\phi_{n=0} = e^{-\sqrt{2m}\int \Delta W_{n=0} dz} = (1 - e^{-\alpha r})^\delta e^{-\sqrt{2m}Br} \quad . \tag{11}$$

Thus, the full solutions corresponding the total potential $-U_0/(1 - e^{\alpha r}) + (S_0^2 - V_0^2)/(1 - e^{-\alpha r})^2$ are

$$E_{n=0}^2 - m^2 = \varepsilon_{n=0} + \Delta\varepsilon_{n=0} = -\frac{1}{8m\alpha^2}\left[\frac{2mU_0}{\delta+1} - \alpha^2\right]^2 \quad ,$$

$$\psi_{n=0} = \chi_{n=0}\phi_{n=0} = (1 - e^{-\alpha r})^{\delta+1} e^{-\left[\frac{mU_0}{\alpha(\delta+1)} + \frac{\alpha}{2}\right]r} \quad . \tag{12}$$

The results agree with [7]. The justifaction of the scenario used in terms of the findings above can also be easily observed if one starts directly from the K-G equation and use the introduced form of $W_{n=0}^{SUSY}$ in a Riccati equation similar to (5) but for the whole potential. However, such a treatment is not so practical due to the screening of the relativistic contributions in the calculation results.

Though we have considered only the ground state solutions here, the extension of the prescription used to the excited states does not cause any problem if the potential in (5) has an algebraic solution. For the clarification of this point, the reader is referred to [1]. Neither the Hulthen potential [8] nor the effective Hulthen like potential appeared here are shape invariant [6], unlike the wrong consideration in the recent analysis of the same problem [5]. Therefore, an algebraic expression for the whole spectrum of the total potential is not available.

Furthermore, as the use of (10) in (6) reproduces $\delta(\delta+1)\frac{\alpha^2}{2m} = S_0^2 - V_0^2$, one can safely express the parameter $\delta$, related to the relativistic contributions through strong interactions in case the scaler potential is larger than the vector potential, in the explicit form

$$\delta = -\frac{1}{2} + \sqrt{\frac{2m}{\alpha^2}(S_0^2 - V_0^2) + \frac{1}{4}} \quad , \tag{13}$$

which supports the earlier work in [7] and physically interesting discussion therein regarding the relation between the reasonable solutions and the strengths of vector/tensor potentials through the parameter $\delta$.

As the second illustration, we focus on the recently investigated [3] mixed perturbed Coulomb like scalar and vector potentials,

$$S(r) = \frac{S_0}{r} + S_1 r + S_2 r^2 \quad , \quad V(r) = \frac{V_0}{r} + V_1 r + V_2 r^2 \quad . \tag{14}$$

Although this problem has been well discussed in the literature with the consideration of exact solvability depending on the potential parameters, there is an alternative case stayed behind the study in [3], which is one of the subject of this section. Seconly, and more significantly, the theoretical consideration here proposes a scheme for a systematic treatment of the relativistic effects if the corresponding equation, Eq. (4) or Eq. (6), is not analytically solvable, whereas the work in [3] lacks of such flexibility.

By the use of Eq. (5) and considering the whole discussion in the Letter, one finds the corresponding solution in a closed algebraic form for the potential, $2[(mS_0 + E_n V_0/r) + (mS_1 + E_n V_1)r + (mS_2 + E_n V_2)r^2]$, in the non-relativistic region where $S = \pm V$, with the choice

$$W_{n=0} = \sqrt{\frac{m}{2}}a - \frac{1}{\sqrt{2m}r} + \sqrt{c}r \quad , \tag{15}$$

in which $a = -2(mS_0 + E_n V_0)$ and $c = 2(mS_2 + E_n V_2)$. This choice, with the natural restriction on the potential parameters such that

$$mS_1 + E_n V_1 = -(mS_0 + E_n V_0)\sqrt{4m(mS_2 + E_n V_2)} \quad , \tag{16}$$

reveals the binding energy at the non-relativistic limit as

$$\varepsilon_n = -(b^2/4c) + \sqrt{c}(2n+3)/\sqrt{2m} = -2m(mS_0 + E_n V_0)^2 + (2n+3)\sqrt{(S_2 + E_n V_2/m)} \tag{17}$$

where $b = 2(mS_1 + E_n V_1)$. The wave function in this domain can readily be calculated in the light of (5). A detailed study of a similar problem in arbitrary dimensions can be found in [9]. We also refer to the related references therein for the complicated relationship between the potential parameters and the radial quantum number $(n = 0,1,2...)$.

In spite of the shape invariance character of the potential in the non-relativistic limit discussed above, having a closed algebraic form for the whole spectrum, the inclusion of the relativistic effects $(S^2 - V^2 \succ 0)$ turns the total potential into the quasi-exactly solvable case [10], which is indeed interesting from the physical point of view. In the contrary, it is reminded that for instance the usual exponential potential has no analytical solution at the non-relativistic domain but the relativistic contribution transforms it an exactly solvable Morse like potential. At this point however, we suggest an alternative scenario for the approximate calculation of relativistic contributions for any quantum state, as long as Eq. (5) is analytically solvable as in the present case. Namely, if one expands Eq. (6), see Ref. [1] for the details, up to e.g. third order then obtains

$$2W_n \Delta W_{n1} - \frac{\Delta W'_{n1}}{\sqrt{2m}} = \Delta V_1 - \Delta \varepsilon_{n1} \quad , \tag{18}$$

$$\Delta W_{n1}^2 + 2W_n \Delta W_{n2} - \frac{\Delta W'_{n2}}{\sqrt{2m}} = \Delta V_2 - \Delta \varepsilon_{n2} \quad , \tag{19}$$

$$2(W_n \Delta W_{n3} + \Delta W_{n1} \Delta W_{n2}) - \frac{\Delta W'_{n3}}{\sqrt{2m}} = \Delta V_3 - \Delta \varepsilon_{n3} \quad , \tag{20}$$

keeping in mind that

$$\Delta V(r;\lambda) = \sum_{k=1}^{\infty} \lambda^k \Delta V_k(r), \quad \Delta W_n(r;\lambda) = \sum_{k=1}^{\infty} \lambda^k \Delta W_{nk}(r), \quad \Delta \varepsilon_n(\lambda) = \sum_{k=1}^{\infty} \lambda^k \Delta \varepsilon_{nk} \tag{21}$$

where $\lambda$ and $k$ denote the perturbation parameter and perturbation order, respectively. It should be remarked that as the system is algebraicly solvable in the nonrelativistic domain, which means that the corresponding wave functions for the all states are known explicitly, one can easily define $W_n = - \chi'_n / \sqrt{2m} \chi_n$ to be used through Eqs. (18-20).

To proceed further, considering the perturbation potential shifted by $2(S_0 S_1 - V_0 V_1)$ because of the relativistic effects,

$$\Delta V = S^2 - V^2 = \frac{(S_0^2 - V_0^2)}{r^2} + 2(S_0 S_2 - V_0 V_2)r + (S_1^2 - V_1^2)r^2 + 2(S_1 S_2 - V_1 V_2)r^3 + (S_2^2 - V_2^2)r^4,$$

(22)

one needs to chose proper $\Delta W_{nk}$ values to satisfy equations at successive perturbation orders such as (18-20) which lead to the approximate energy $\Delta \varepsilon_n = \sum_j \Delta \varepsilon_{nj}$ and wave function $\phi_n = e^{-\sqrt{2m} \int \sum_k \Delta W_{nk} dz}$ values to obtain the modified relativistic extension, $E_n^2 - m^2 = \varepsilon_n + \Delta \varepsilon_n$ and $\psi_n = \chi_n \phi_n$, of the results in the non-relativistic domain. Though, this formalism does not seem so practical, in particular for the system under consideration due to the quite complicated relationship between the potential parameters in higher quantum states, it could be easy for the other physical systems [1] and may work efficiently.

The present systematic study obviously recovers a number of earlier results for many different potentials in a natural unified way and also leads to new findings. The idea put forward in this Letter would be used to explore a great number of relativistic systems and can be also extended to the case of the Dirac equation. Furthermore, the possibility of getting approximate solutions from relativistic (non-relativistic) ones, if the relativistic (non-relativistic) solutions have an algebraic form, should definitely be investigated more profoundly. With this connection, the search of relativistic equations with position dependent mass [11] would be interesting. Along these lines the works are in progress, which will be deferred to another publication.